\documentclass[conference]{IEEEtran}

\makeatletter
\def\ps@IEEEtitlepagestyle{%
  \def\@oddfoot{\mycopyrightnotice}%
  \def\@evenfoot{}%
}
\def\mycopyrightnotice{%
  {\footnotesize XXX-X-XXXX-XXXX-X/XX/\$XX.00~\copyright~2026 IEEE\hfill}
  \gdef\mycopyrightnotice{}
}

\usepackage{blindtext}
\usepackage{eso-pic}
\IEEEoverridecommandlockouts
\usepackage{cite}
\usepackage{amsmath,amssymb,amsfonts}
\usepackage{algorithmic}
\usepackage{graphicx}
\usepackage{textcomp}
\usepackage{xcolor}
\usepackage{orcidlink} 
\def\BibTeX{{\rm B\kern-.05em{\sc i\kern-.025em b}\kern-.08em
    T\kern-.1667em\lower.7ex\hbox{E}\kern-.125emX}}
    
\usepackage{eso-pic}

\usepackage{balance}    
    
\begin{document}
\title{\vspace*{1cm} Operational Collapse Region in Repeaterless Loss-Dephasing Quantum Channels
\thanks{Scientific and Technological
Research Council of Turkey (T\"{U}B\.{I}TAK, Grant No.\ 125F473).}
}

\author{\IEEEauthorblockN{1\textsuperscript{st} Ufuk Korkmaz~\orcidlink{0000-0001-5836-5262}}
\IEEEauthorblockA{\textit{Informatics Institute} \\
\textit{Istanbul Technical University}\\
\textit{Qready Quantum Technologies}\\
\textit{ITU ARI Teknokent}\\
\.{I}stanbul, T\"{u}rkiye \\
ufukkorkmaz@itu.edu.tr}
\and
\IEEEauthorblockN{2\textsuperscript{nd} S. Elham {Mousavigharalari}~\orcidlink{0009-0004-4840-1388}}
\IEEEauthorblockA{\textit{Informatics Institute} \\
\textit{Istanbul Technical University}\\
\.{I}stanbul, T\"{u}rkiye \\
mousavie@itu.edu.tr}
\and
\IEEEauthorblockN{3\textsuperscript{rd} Deniz Türkpençe~\orcidlink{0000-0002-5182-374X}}
\IEEEauthorblockA{\textit{Informatics Institute} \\
\textit{Istanbul Technical University}\\
\textit{Qready Quantum Technologies}\\
\textit{ITU ARI Teknokent}\\
\.{I}stanbul, T\"{u}rkiye \\
dturkpence@itu.edu.tr}

}

\maketitle
\begin{abstract}
The distribution of entangled photon pairs over standard optical
fiber is a fundamental requirement for the realization of the
quantum internet. However, real-world deployment is severely
bottlenecked by the interplay of amplitude damping (photon loss)
and phase noise (birefringence). In this paper, we numerically
investigate the degradation of dual-rail polarization entanglement
in telecom C-band fiber links. We demonstrate a critical disparity
between the physical survival of quantum correlations and their
practical utility in standard communication protocols. By evaluating
the unconditional logarithmic negativity against the post-selected
teleportation fidelity, we identify a distinct ``operational collapse
region''---a distance window where the channel retains true quantum
entanglement, yet standard coincidence-based detection architectures
fail to provide any advantage over classical strategies. Furthermore,
we reveal that the width of this inaccessible region exhibits a
non-monotonic dependence on the phase noise rate, implying that
simply minimizing fiber dephasing does not necessarily optimize the
operational efficiency of the network. These findings provide vital
guidelines for the design of practical quantum communication links.
\end{abstract}


\section{Introduction}
The transition of quantum communication from laboratory experiments
to deployed optical fiber networks introduces severe engineering
challenges. Central to protocols such as Quantum Key Distribution
(QKD) and quantum teleportation~\cite{gisin_quantum_2002,
scarani_security_2009} is the faithful distribution of entangled
photon pairs. Telecom fibers inherently impose two dominant noise
mechanisms: amplitude damping due to material absorption and
scattering, and pure dephasing caused by polarization mode
dispersion and environmental phase fluctuations~\cite{gisin_quantum_2002,agrawal_fiber_2012}.

In evaluating the performance of a quantum network link, an
essential distinction must be made between the theoretical
existence of quantum correlations and their practical
exploitability~\cite{pirandola_fundamental_2017,mele_quantum_2024}.
Traditional analyses often characterize channel performance
through asymptotic capacity bounds. However, from a network
engineering perspective, an entangled resource becomes
operationally obsolete when its fidelity drops below the
threshold required to outperform classical measure-and-prepare
strategies~\cite{horodecki_general_1999}. The gap between the
extinction of entanglement and the collapse of protocol
usefulness has not been systematically mapped in the telecom
fiber regime~\cite{ecker_overcoming_2019}.

In this work, we present an open-system numerical analysis of
dual-rail polarization-entangled photons transmitted through a
symmetric loss-dephasing telecom channel. All numerical
simulations are performed using the QuTiP
framework~\cite{johansson_qutip_2013}. We contrast an
unconditional entanglement metric---which encompasses the
complete physical state including photon loss events---with a
conditional operational metric based on coincidence
post-selection. Our simulations reveal a distinct ``dead zone''
where standard communication hardware becomes blind to the
residual entanglement in the fiber, and show that the spatial
width of this operational gap varies non-monotonically with
the environmental phase noise.

Targeting the telecom C-band directly aligns our framework with 
the physical constraints of contemporary Metropolitan Area Networks (MANs). 
While laboratory environments can employ advanced active stabilization 
to artificially nullify phase fluctuations, deployed metropolitan fibers 
are continuously subjected to heterogeneous environmental stresses. 
Recent field demonstrations of such metropolitan networks 
have achieved high-fidelity teleportation over long-distance standard 
optical fibers and hybrid links by utilizing dissimilar quantum dots 
and advanced synchronization techniques~\cite{laneve_quantum_2025}.

However, since direct repeaterless transmission in these 
networks inevitably suffers from simultaneous photon loss and dephasing, 
our quantified operational collapse region dictates the theoretical limit 
for passive dual-rail polarization channels. To mitigate these fundamental 
limitations, recent advancements emphasize the necessity of dynamic 
entanglement routing and memory-assisted architectures~\cite{abane_entanglement_2025}. 
Integrating solid-state quantum memories---particularly Erbium-ion ensembles 
that operate natively in the telecommunication C-band~\cite{an_quantum_2025}---can 
serve as a critical buffer. By temporarily storing the quantum state before 
the entanglement is irreversibly degraded, hybrid receiver architectures 
can effectively bypass the dead zones characterized in our models~\cite{an_quantum_2025, laneve_quantum_2025}.

Therefore, identifying the exact boundaries of the operational collapse 
region not only provides a vital benchmark for scheduling 
and resource allocation in near-term quantum infrastructure, but 
also establishes a stringent operational threshold for triggering entanglement 
swapping and memory retrieval in future global quantum internets.

\section{System Model and Metrics}
To rigorously quantify the boundaries of the operational collapse region, we construct a comprehensive open-system model of the fiber link. This section outlines the mathematical framework underpinning our numerical simulations. First, we define the physical encoding of the dual-rail quantum state and describe its non-unitary evolution under simultaneous photon loss and environmental dephasing mechanisms. Subsequently, we introduce the distinct theoretical and operational performance metrics required to evaluate both the state-level entanglement survival and its hardware-level utility.

\subsection{Encoding and Channel Dynamics}

We encode the qubit states in the dual-rail polarization basis:
$|H\rangle \equiv |1\rangle_H \otimes |0\rangle_V$ and
$|V\rangle \equiv |0\rangle_H \otimes |1\rangle_V$,
where $|n\rangle_\lambda$ denotes the $n$-photon Fock state of
mode $\lambda \in \{H,V\}$. This places the state naturally
within the bosonic Fock space and allows the noise operators to
be written directly in terms of mode operators. The initial
resource is the dual-rail Bell state
$|\Phi^+\rangle = (|HH\rangle + |VV\rangle)/\sqrt{2}$.

As the photons propagate through the fiber, the evolution of the
full $16$-dimensional bipartite density matrix $\rho(d)$ over
distance $d$ is governed by the distance-parameterized Lindblad
master equation~\cite{lindblad_generators_1976}
\begin{equation}
  \frac{d\rho}{dz}
  = \sum_k \!\left(
      L_k \rho L_k^\dagger
      - \tfrac{1}{2} L_k^\dagger L_k \rho
      - \tfrac{1}{2} \rho L_k^\dagger L_k
    \right),
  \label{eq:lindblad}
\end{equation}
where $z$ denotes the propagation distance along the fiber.
We incorporate two independent local noise operators for each
arm: an amplitude damping operator
$L_{\mathrm{loss}} = \sqrt{\kappa}\, a$
representing photon attenuation, and a dephasing operator
$L_{\mathrm{deph}} = \sqrt{\gamma_\phi}\, \sigma_z$
representing polarization phase diffusion. The loss rate is
related to the fiber attenuation via
$\kappa = \tfrac{\ln 10}{10}\,\alpha$.

\subsection{Performance Metrics and Thresholds}

To capture the disparity between state-level physics and
hardware-level utility, we track two distinct metrics.

\paragraph{Unconditional logarithmic negativity ($E_{\mathcal{N}}$).}
This metric is evaluated on the full $16$-dimensional density
matrix $\rho(d)$, including vacuum and partial-loss sectors,
without post-selection. It therefore probes whether the physical
channel output retains quantum
correlations~\cite{vidal_computable_2002}. Because
$E_{\mathcal{N}}$ decays asymptotically, we define the
entanglement extinction distance $d_E$ via a practical threshold:
$E_{\mathcal{N}}(d_E) = \varepsilon_{E_N}$.
Certifying entanglement of magnitude $\varepsilon$ requires
$\mathcal{O}(\varepsilon^{-2})$ coincidence
measurements~\cite{flammia_direct_2011}; at metropolitan fiber
coincidence rates, resolving values below
$\varepsilon_{E_N} = 10^{-4}$ would demand integration times
inconsistent with fiber link
stability~\cite{friis_entanglement_2019}.
We therefore adopt $\varepsilon_{E_N} = 10^{-4}$ throughout.

\paragraph{Post-selected teleportation fidelity ($F_{\mathrm{tele}}$).}
Standard coincidence-based hardware post-selects on the
successful arrival of both photons. On this surviving subspace,
we evaluate the average teleportation fidelity $F_{\mathrm{tele}}$
via the Horodecki relation~\cite{horodecki_general_1999},
which gives $F_{\mathrm{tele}} = (2f_{\max}+1)/3$, where
$f_{\max}$ is the maximal Bell-state overlap of the
post-selected state. A distributed state ceases to provide a
quantum advantage when $F_{\mathrm{tele}} \le 2/3$.
We define the operational collapse distance $d_F$ by
$F_{\mathrm{tele}}(d_F) = 2/3 + \delta_F$ with
$\delta_F = 10^{-3}$, to avoid asymptotic numerical artifacts.
The post-selection success probability
$p_{\mathrm{succ}}(d) = \mathrm{Tr}[\Pi\,\rho(d)]$,
where $\Pi$ projects onto the logical two-qubit subspace,
is reported alongside $F_{\mathrm{tele}}$ to prevent
misreading of conditional quality as unconditional performance.

The operational collapse region is then
\begin{equation}
  \mathcal{R}_{\mathrm{dead}}
  = \bigl\{z \;\big|\;
    E_{\mathcal{N}}(z) > \varepsilon_{E_N}
    \;\text{ and }\;
    F_{\mathrm{tele}}(z) \le \tfrac{2}{3} + \delta_F
  \bigr\},
  \label{eq:Rdead}
\end{equation}
with width $\Delta d = d_E - d_F$.

\section{Numerical Results}
In this section, we present the outcomes of our open-system simulations, executed via the QuTiP framework, to map the degradation of polarization entanglement over standard telecom fibers. We first analyze the distance-resolved dynamics of the dual-rail Bell state under baseline C-band parameters to clearly demonstrate the emergence of the operational collapse region. Subsequently, we perform a parameter sweep to investigate the sensitivity of this dead zone's width to varying rates of environmental phase noise, revealing the non-monotonic relationship that governs the link's operational efficiency.

\subsection{The Operational Collapse Region}

We first establish the baseline dynamics with parameters
calibrated for standard single-mode fibers in the telecom
C-band: $\alpha = 0.2$~dB/km ($\kappa \approx 0.046$~km$^{-1}$)
and $\gamma_\phi = 0.08$~km$^{-1}$.

\begin{figure}[t]
  \centering
  \includegraphics[width=\columnwidth]{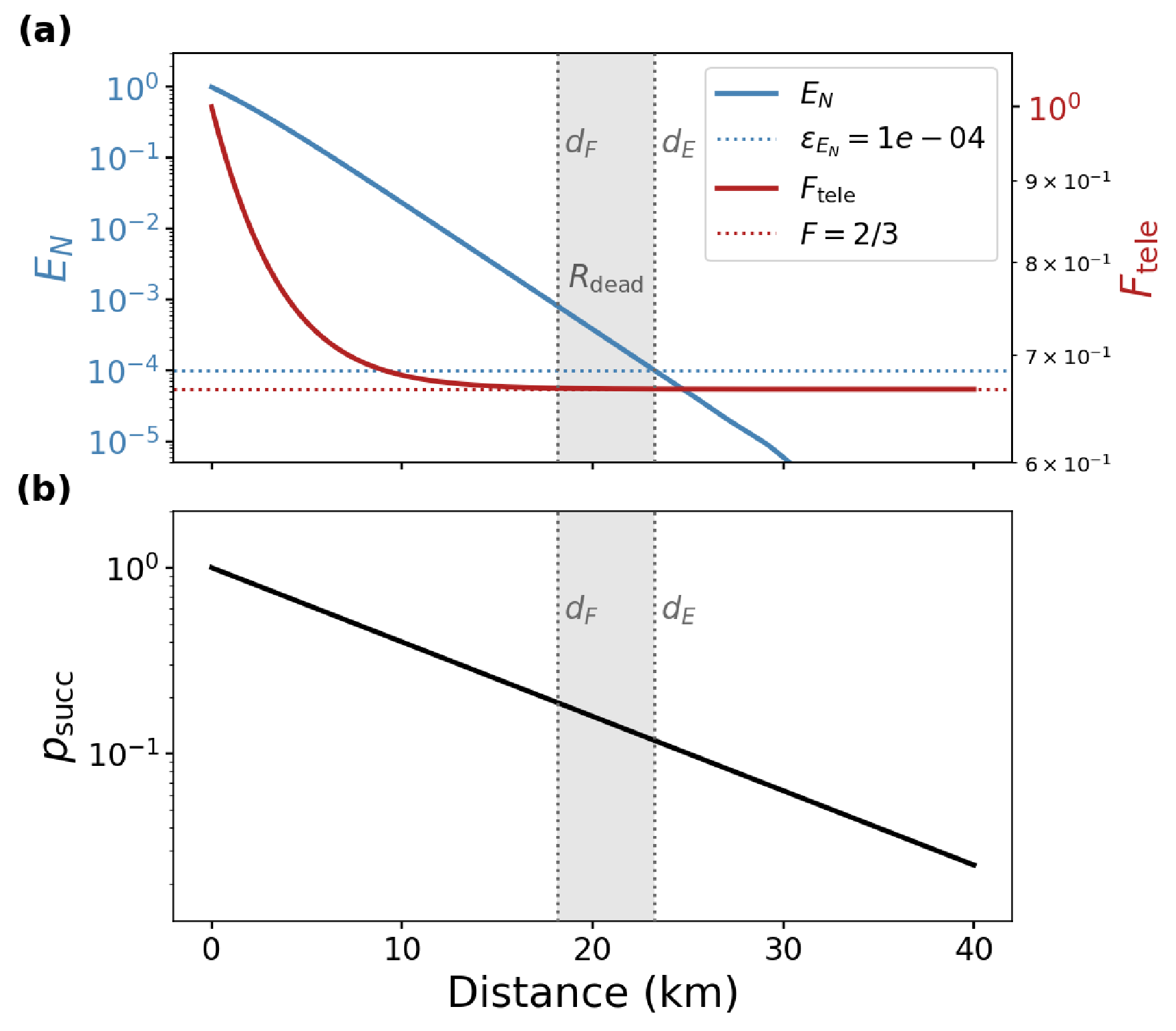}
  \caption{Distance-resolved dynamics of a dual-rail Bell state
  in a telecom C-band channel
  ($\alpha = 0.2$~dB/km, $\gamma_\phi = 0.08$~km$^{-1}$).
  Left axis (blue): unconditional logarithmic negativity
  $E_{\mathcal{N}}$; dotted line marks
  $\varepsilon_{E_N} = 10^{-4}$.
  Right axis (red): post-selected teleportation fidelity
  $F_{\mathrm{tele}}$; dotted line marks the classical
  benchmark $F = 2/3$.
  The grey shaded band is $\mathcal{R}_{\mathrm{dead}}$
  ($\Delta d \approx 5.1$~km), bounded by
  $d_F \approx 18.2$~km and $d_E \approx 23.2$~km.
  Panel~(b): post-selection success probability
  $p_{\mathrm{succ}}(d)$.}
  \label{fig:baseline}
\end{figure}

As shown in Fig.~\ref{fig:baseline}, the two metrics degrade
on clearly distinct spatial scales. The post-selected
teleportation fidelity crosses the classical threshold at
$d_F \approx 18.2$~km, at which point the distributed state
provides no teleportation advantage over classical strategies.
The unconditional $E_{\mathcal{N}}$, however, confirms that
the global state remains entangled until $d_E \approx 23.2$~km,
yielding $\mathcal{R}_{\mathrm{dead}}$ with
$\Delta d \approx 5.1$~km.

The physical origin of this gap is transparent. Photon loss
transfers population to the vacuum sector, suppressing
$E_{\mathcal{N}}$ evaluated over the full state. The
post-selected $F_{\mathrm{tele}}$, by contrast, is evaluated
only on surviving photon pairs and is insensitive to loss as
such; it is degraded primarily by dephasing, which erodes
the polarization coherence of the surviving subspace.
The two noise processes therefore affect the two metrics
on different distance scales, opening a window in which
the fiber continues to carry quantum correlations while
standard coincidence-based protocols are unable to exploit them.

Panel~(b) of Fig.~\ref{fig:baseline} shows that
$p_{\mathrm{succ}}$ falls from unity at $d=0$ to approximately
$0.063$ at $d_F$ and $0.034$ at $d_E$.
This context is important: the high conditional fidelity of
the surviving pairs at large distances does not indicate
good channel performance, but rather that the small fraction
of photons that avoid loss retain relatively clean polarization
coherence. Within $\mathcal{R}_{\mathrm{dead}}$, the resource
is physically present but operationally inaccessible to
standard hardware.

\subsection{Impact of Phase Noise on Gap Width}

We next examine how $\mathcal{R}_{\mathrm{dead}}$ changes
as the dephasing rate is varied at fixed attenuation
$\alpha = 0.2$~dB/km. Four representative values of
$\gamma_\phi$ are selected, ranging from a
dephasing-dominated regime to a loss-dominated one.
The thresholds $\varepsilon_{E_N}$ and $\delta_F$ are
unchanged throughout, ensuring that differences in
$\Delta d$ arise from the channel physics alone.

\begin{figure}[t]
  \centering
  \includegraphics[width=\columnwidth]{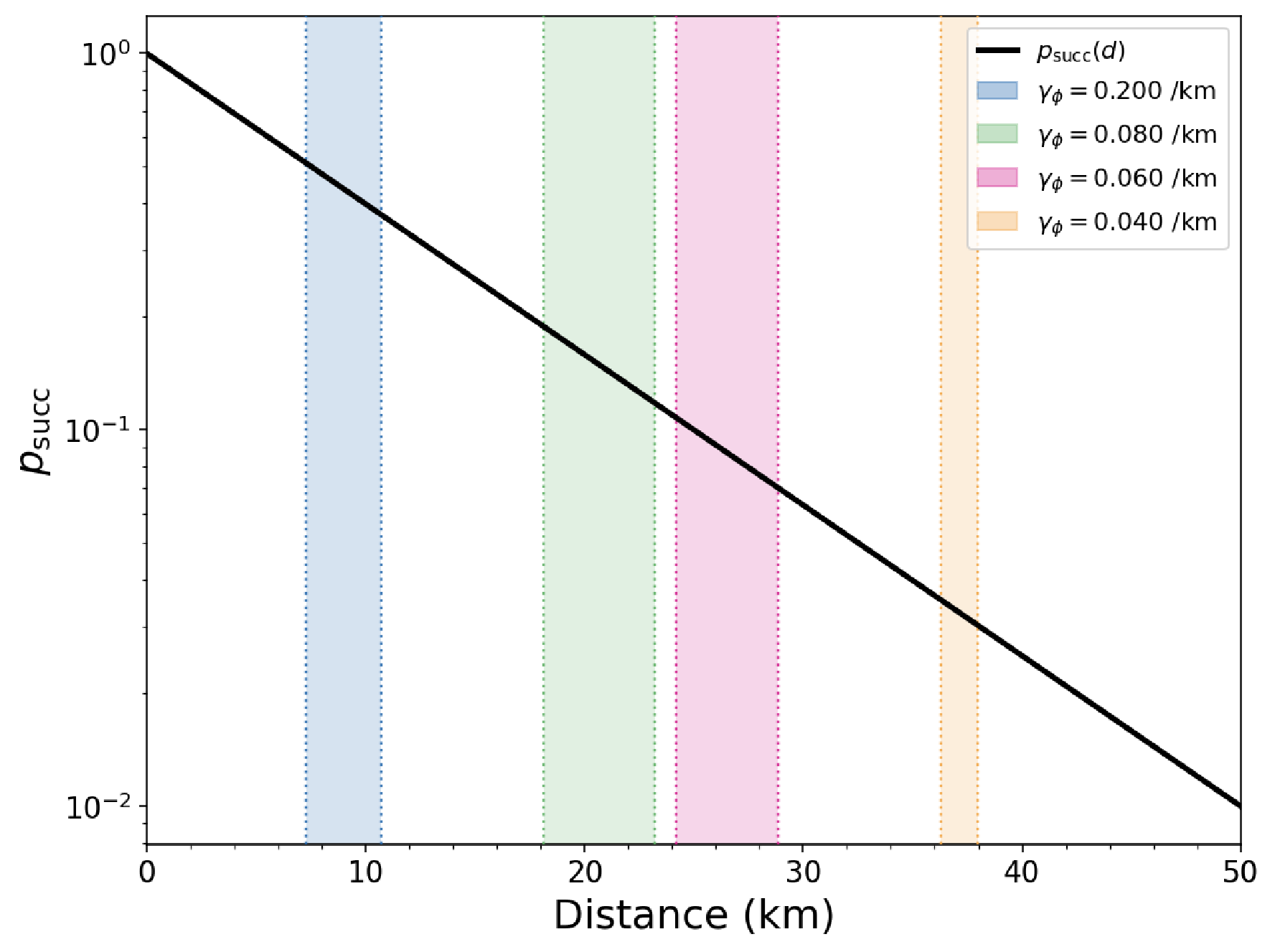}
  \caption{Post-selection success probability $p_{\mathrm{succ}}(d)$
  (black curve) for a symmetric telecom channel with
  $\alpha = 0.2$~dB/km ($\kappa \approx 0.046$~km$^{-1}$).
  Colored bands indicate $\mathcal{R}_{\mathrm{dead}}$ for
  four values of the dephasing rate:
  $\gamma_\phi \in \{0.20,\, 0.08,\, 0.06,\, 0.04\}$~km$^{-1}$.
  Each band spans $[d_F,\, d_E]$, where $d_F$ is defined by
  $F_{\mathrm{tele}}(d_F) = 2/3 + \delta_F$
  ($\delta_F = 10^{-3}$) and $d_E$ by
  $E_{\mathcal{N}}(d_E) = \varepsilon_{E_N}$
  ($\varepsilon_{E_N} = 10^{-4}$).
  The gap width $\Delta d$ reaches a maximum of $5.1$~km
  at $\gamma_\phi = 0.08$~km$^{-1}$ and contracts at
  both higher and lower dephasing rates.}
  \label{fig:sweep}
\end{figure}

The results in Fig.~\ref{fig:sweep} reveal a non-monotonic
dependence of $\Delta d$ on $\gamma_\phi$.
At high dephasing ($\gamma_\phi = 0.35$~km$^{-1}$), phase
noise destroys both metrics at short distances and the gap
is narrow ($\Delta d = 2.3$~km, $d_F = 4.2$~km,
$d_E = 6.4$~km).
As $\gamma_\phi$ decreases, the gap widens, reaching a maximum
of $\Delta d = 5.1$~km at $\gamma_\phi = 0.08$~km$^{-1}$
($d_F = 18.2$~km, $d_E = 23.2$~km).
Further reduction of dephasing does not continue to widen
the gap: at $\gamma_\phi = 0.06$~km$^{-1}$,
$\Delta d = 4.6$~km, and at $\gamma_\phi = 0.04$~km$^{-1}$,
$\Delta d = 1.7$~km ($d_F = 36.3$~km, $d_E = 38.0$~km).

The mechanism is as follows. Reducing $\gamma_\phi$ postpones
the collapse of $F_{\mathrm{tele}}$ (pushing $d_F$ to larger
distances), but it has comparatively little effect on
$d_E$, which is also constrained by loss. As $\gamma_\phi$
becomes very small, $d_F$ continues to grow while $d_E$
approaches a loss-determined ceiling, causing the gap to
contract. The maximum $\Delta d$ therefore occurs at an
intermediate ratio of loss to dephasing, and this ratio
is determined entirely by the structure of the two competing
decay mechanisms rather than by any fine-tuning of parameters.
This behavior is structurally robust: the non-monotonic
trend persists across the full range of simulated values and
is insensitive to moderate changes in the operational
thresholds.

\section{Conclusion}

We have shown that the operational lifespan of a dual-rail
quantum link is in general shorter than the physical lifespan
of the entanglement it carries. The operational collapse
region $\mathcal{R}_{\mathrm{dead}}$---in which the channel
remains entangled at the state level while standard protocols
fail to outperform classical strategies---exists over a finite
distance window whose width depends non-monotonically on the
dephasing rate. The maximum gap occurs at an intermediate ratio
of loss to dephasing, implying that standard coincidence-based
detection introduces a systematic discrepancy between physical
and operational entanglement lifetimes in realistic fiber
links, and that this discrepancy is largest not at the
noisiest extreme but at a specific intermediate noise regime.
These results suggest that hardware-level protocol design should
account for the interplay between loss and dephasing rather
than treating them as independently optimizable parameters.

From a broader technological perspective, the existence of this 
dead zone underscores a substantial reservoir of wasted quantum 
potential. The residual unconditional entanglement in the high-loss 
regime is largely dominated by superpositions involving the vacuum state. 
While traditional avalanche photodiodes or superconducting nanowire 
detectors discard these events via post-selection, future receiver 
architectures—such as continuous-variable homodyne systems and memory-assisted hybrid interfaces~\cite{an_quantum_2025}—might bypass 
this bottleneck. Furthermore, actively integrating our characterized operational thresholds into dynamic entanglement routing algorithms will be crucial for maintaining end-to-end fidelity in near-term metropolitan networks~\cite{abane_entanglement_2025, laneve_quantum_2025}. Until such advanced interfaces become commercially viable, 
quantifying the operational gap remains an indispensable step for 
validating the true capacity of repeaterless networks and establishing the precise hardware requirements for future global quantum internets.

\section*{Acknowledgment}

This work was supported by the Scientific and Technological
Research Council of Turkey (T\"{U}B\.{I}TAK, Grant No.\ 125F473).
We also acknowledge the facilities and technical support provided
by the Informatics Institute of Istanbul Technical University
and Qready Quantum Technologies and Consulting Joint Stock Company.

\balance


\begin{thebibliography}{10}
\providecommand{\url}[1]{#1}
\csname url@samestyle\endcsname
\providecommand{\newblock}{\relax}
\providecommand{\bibinfo}[2]{#2}
\providecommand{\BIBentrySTDinterwordspacing}{\spaceskip=0pt\relax}
\providecommand{\BIBentryALTinterwordstretchfactor}{4}
\providecommand{\BIBentryALTinterwordspacing}{\spaceskip=\fontdimen2\font plus
\BIBentryALTinterwordstretchfactor\fontdimen3\font minus
  \fontdimen4\font\relax}
\providecommand{\BIBforeignlanguage}[2]{{%
\expandafter\ifx\csname l@#1\endcsname\relax
\typeout{** WARNING: IEEEtran.bst: No hyphenation pattern has been}%
\typeout{** loaded for the language `#1'. Using the pattern for}%
\typeout{** the default language instead.}%
\else
\language=\csname l@#1\endcsname
\fi
#2}}
\providecommand{\BIBdecl}{\relax}
\BIBdecl

\bibitem{gisin_quantum_2002}
N.~Gisin, G.~Ribordy, W.~Tittel, and H.~Zbinden, ``Quantum cryptography,''
  \emph{Reviews of Modern Physics}, vol.~74, no.~1, pp. 145--195, 2002.

\bibitem{scarani_security_2009}
V.~Scarani, H.~Bechmann-Pasquinucci, N.~J. Cerf, M.~Dušek, N.~Lütkenhaus, and
  M.~Peev, ``The security of practical quantum key distribution,''
  \emph{Reviews of Modern Physics}, vol.~81, no.~3, pp. 1301--1350, 2009.

\bibitem{agrawal_fiber_2012}
G.~P. Agrawal, \emph{Fiber-Optic Communication Systems}.\hskip 1em plus 0.5em
  minus 0.4em\relax Johan Wiley \& Sons, 2012.

\bibitem{pirandola_fundamental_2017}
S.~Pirandola, R.~Laurenza, C.~Ottaviani, and L.~Banchi, ``Fundamental limits of
  repeaterless quantum communications,'' \emph{Nature Communications}, vol.~8,
  p. 15043, 2017.

\bibitem{mele_quantum_2024}
F.~A. Mele, F.~Salek, V.~Giovannetti, and L.~Lami, ``Quantum communication on
  the bosonic loss-dephasing channel,'' \emph{Physical Review A}, vol. 110, p.
  012460, 2024.

\bibitem{horodecki_general_1999}
M.~Horodecki, P.~Horodecki, and R.~Horodecki, ``General teleportation channel,
  singlet fraction, and quasidistillation,'' \emph{Physical Review A}, vol.~60,
  no.~3, pp. 1888--1898, 1999.

\bibitem{ecker_overcoming_2019}
S.~Ecker, F.~Bouchard, L.~Bulla, F.~Brandt, O.~Kohout, F.~Steinlechner,
  R.~Fickler, M.~Malik, Y.~Guryanova, R.~Ursin \emph{et~al.}, ``Overcoming
  noise in entanglement distribution,'' \emph{Physical Review X}, vol.~9,
  no.~4, p. 041042, 2019.

\bibitem{johansson_qutip_2013}
J.~R. Johansson, P.~D. Nation, and F.~Nori, ``{QuTiP} 2: {A} {Python} framework
  for the dynamics of open quantum systems,'' \emph{Computer Physics
  Communications}, vol. 184, no.~4, pp. 1234--1240, 2013.

\bibitem{laneve_quantum_2025}
A.~Laneve, G.~Ronco, M.~Beccaceci, P.~Barigelli, F.~Salusti,
  N.~Claro-Rodriguez, G.~De~Pascalis, A.~Suprano, L.~Chiaudano, E.~Schöll,
  L.~Hanschke, T.~M. Krieger, Q.~Buchinger, S.~F. Covre~da Silva, J.~Neuwirth,
  S.~Stroj, S.~Höfling, T.~Huber-Loyola, M.~A. Usuga~Castaneda, G.~Carvacho,
  N.~Spagnolo, M.~B. Rota, F.~Basso~Basset, A.~Rastelli, F.~Sciarrino, K.~D.
  Jöns, and R.~Trotta, ``Quantum teleportation with dissimilar quantum dots
  over a hybrid quantum network,'' \emph{Nature Communications}, vol.~16,
  no.~1, p. 10028, Nov. 2025.

\bibitem{abane_entanglement_2025}
A.~Abane, M.~Cubeddu, V.~S. Mai, and A.~Battou, ``Entanglement {Routing} in
  {Quantum} {Networks}: {A} {Comprehensive} {Survey},'' \emph{IEEE Transactions
  on Quantum Engineering}, vol.~6, pp. 1--39, 2025.

\bibitem{an_quantum_2025}
Y.-Y. An, Q.~He, W.~Xue, M.-H. Jiang, C.~Yang, Y.-Q. Lu, S.~Zhu, and X.-S. Ma,
  ``Quantum {Teleportation} from {Telecom} {Photons} to {Erbium}-{Ion}
  {Ensembles},'' \emph{Physical Review Letters}, vol. 135, no.~1, p. 010804,
  Jul. 2025.

\bibitem{lindblad_generators_1976}
G.~Lindblad, ``\BIBforeignlanguage{en}{On the generators of quantum dynamical
  semigroups},'' \emph{\BIBforeignlanguage{en}{Communications in Mathematical
  Physics}}, vol.~48, no.~2, pp. 119--130, 1976.

\bibitem{vidal_computable_2002}
G.~Vidal and R.~F. Werner, ``Computable measure of entanglement,''
  \emph{Physical Review A}, vol.~65, no.~3, p. 032314, 2002.

\bibitem{flammia_direct_2011}
S.~T. Flammia and Y.-K. Liu, ``Direct fidelity estimation from few {Pauli}
  measurements,'' \emph{Physical Review Letters}, vol. 106, p. 230501, 2011.

\bibitem{friis_entanglement_2019}
N.~Friis, G.~Vitagliano, M.~Malik, and M.~Huber, ``Entanglement certification
  from theory to experiment,'' \emph{Nature Reviews Physics}, vol.~1, pp.
  72--87, 2019.

\end{thebibliography}

\end{document}